\documentclass[reprint,nofootinbib,superscriptaddress,amsmath,amssymb,aps]{revtex4-1}
\usepackage{graphicx}
\usepackage{rotating}
\usepackage{dcolumn}
\usepackage{bm}
\usepackage{color}
\usepackage{mathptmx, textcomp}
\usepackage[latin1]{inputenc}
\usepackage{braket}
\usepackage[colorlinks,linkcolor=magenta,anchorcolor=cyan,citecolor=blue]{hyperref}
\usepackage[multiple]{footmisc}
\newcommand{\fig}[1]{Fig.\ref{#1}}
\def\be{\begin{equation}}
\def\ee{\end{equation}}
\def\ba{\begin{eqnarray}}
\def\ea{\end{eqnarray}}

\def\nn{\nonumber}
\def\lf{\left}
\def\rt{\right}

\newcommand{\eq}[1]{(\ref{#1})}

\def\lf{\left}\def\rt{\right}\def\q{\theta}      \def\p {\pi}   \def\d {\delta} \def\f {\phi}  \def\h {\eta}   \def\l {\lambda}  \def\x {\xi}     \def\pd {\partial}\def\p {\pi}   
\def\Q{\Theta}      \def\S {\Sigma}  \def\F {\Phi}  \def\L {\Lambda}    \def\.{\cdot}
\def\math {\mathcal}
\begin{document}
\title{Examining the weak cosmic censorship conjecture of RN-AdS black holes via the new version of the gedanken experiment}
\author{Xin-Yang Wang}
\email{xinyang\_wang@foxmail.com}
\author{Jie Jiang}
\email{Corresponding author. jiejiang@mail.bnu.edu.cn}
\affiliation{Department of Physics, Beijing Normal University, Beijing, 100875, China}

\date{\today}

\begin{abstract}
Based on the new version of the gedanken experiment proposed by Sorce and Wald, we investigate the weak cosmic censorship conjecture (WCCC) for Reissner-Nordstr\"{o}m-AdS (RN-AdS) black holes under the second-order approximation of the perturbation that comes from the matter fields. First of all, we propose that the cosmological constant is a portion of the matter fields. It means that the cosmological constant can be seen as a variable, and its value will vary with the process of the perturbation. Moreover, we assume that the perturbation satisfies the stability condition, which states that after the perturbation, the spacetime geometry also belongs to the class of RN-AdS solutions. Based on the stability condition and the null energy condition, while using the method of the off-shell variation, the first two order perturbation inequalities are derived. In the two inequalities, the term which contains the thermodynamic pressure and volume is involved firstly. Furthermore, based on the first-order optimal option and the second-order inequality, the WCCC for nearly extremal RN-AdS black holes is examined. It is shown that considering the cosmological constant varying with the perturbation, the WCCC for RN-AdS black holes cannot be violated under the second-order approximation of the matter fields perturbation.
\end{abstract}
\maketitle

\section{Introduction}
The existence of black holes has been predicted by General Relativity. For most black holes, there is a gravitational singularity at the center. Normally, the singularity should be surrounded by the event horizon and hidden inside the black hole. It is because the singularity can make the divergence of the spacetime curvature and destroy the law of causality. If the event horizon vanishes and the naked singularity is exposed to the spacetime, the geometry of the spacetime cannot be mathematically well defined. In order to avoid this situation, Penrose \cite{Penrose:1969pc} proposed the weak cosmic censorship conjecture (WCCC), it states that the singularity should be hidden inside the black hole and the observer at infinity cannot detect any information about it.

Although the WCCC is generally suggested for any black hole, there is not a general procedure to prove the validity of the conjecture. In addition, the validity examination of the WCCC also depends on the selected research method. In 1974, Wald \cite{Wald94} first proposed a gedanken experiment to examine the WCCC for nearly extremal Kerr-Newman (KN) black holes. It states that considering a process that a test particle carrying sufficient charges and angular momentum falls into the black hole, if the particle is absorbed by the black hole, the black hole can overcharge (or overspin) and violate the requirement of the WCCC. However, the results show that the particle cannot be captured by the black hole due to electromagnetic and centrifugal repulsion force. It means that the WCCC for nearly extremal KN black holes cannot be violated at the end of the process. Since then, based on the method proposed by Wald, the WCCC for many kinds of black holes has been examined through considering black holes capture a test particle or a wave packet of a classical field \cite{Cohen:1979zzb, Needham:1980fb, Semiz:1990fm, Bekenstein:1994nx, Semiz:2005gs}. However, this method has an inherent defect because considering the process of the particle dropping into the black hole, the black hole is just treated as a background spacetime. In other words, the interaction between the test particle and the spacetime of the black hole is neglected during the process. Moreover, Hubeny \cite{Hubeny:1998ga} used this method to demonstrate that if one suitably chooses a particle and make it falling into Reissner-Nordstr\"{o}m (RN) black holes, the particle can be captured by the black hole, and the black hole will be overcharged. Therefore, the WCCC can be violated under the process. From this example, it strengthens to illustrate that when ignoring the interaction between the spacetime and test particle, the process of argumentation for the validity of the WCCC has not been given sufficient consideration.

To resolve the inherent in the above method and make the process that the matter drops into the black hole more fully to be considered, the two effects which are the self-force and the back-reaction have been respectively added in the method. Based on the two effects, the WCCC has been proven to hold for Kerr black holes and RN black holes respectively \cite{Barausse:2010ka, Barausse:2011vx, Colleoni:2015ena, Isoyama:2011ea}. Recently, Sorce and Wald \cite{Sorce:2017dst} proposed a new version of the gedanken experiment to examine the WCCC for KN black hole. The matter fields which occur a finite region outside the black hole and sufficiently couple to the spacetime have been considered. The matter fields and the spacetime can be seen as a dynamical system, and the process that the matter fields fall into the black hole can be regarded as a complete dynamic evolution process. Besides, an important assumption is proposed, which states that when the matter fields drop into the black hole, the configuration of spacetime also belongs to the class of KN solutions at the sufficient late times. Based on the assumption, it is shown that under the second-order approximation of the matter fields perturbation, the WCCC cannot be violated. After that, using this method, the WCCC for other kinds of black holes is examined to be valid \cite{An:2017phb, Ge:2017vun, Jiang:2019ige, Jiang:2019vww}. However, for previous studies, only the black holes, which the asymptotic topology is flat, have been investigated, but the situation that the cosmological constant is contained in the black hole spacetime has not been studied. Therefore, we would like to use the method proposed by Sorce and Wald to investigate the WCCC for RN-AdS black holes.

The cosmological constant plays an important role to determine the asymptotic topology of a black hole spacetime. In the theory of Einstein's gravity, the cosmological constant is normally regarded as a constant. However, any constant in the physics theory is not fixed a priori \cite{Cvetic:2010jb}. The renormalization group theory points out that the value of the coupling constant for the Yukawa and the gauge theory depends on the energy scale, which can increase or decrease with the change of the energy scale. The Newton's constant can also change with the expansion of the universe. Therefore, we cannot directly assume that the cosmological constant is a constant. Actually, there are a lot of works to regard the cosmological constant as a dynamic variable, especially in the investigation of the black hole thermodynamics \cite{Teitelboim:1985dp, Padmanabhan:2002sha, Dolan:2010ha, Kubiznak:2012wp, Kubiznak:2016qmn}. For the black hole thermodynamics, when the cosmological constant is considered as a thermodynamic variable, the phase space of thermodynamics can be extended, which is called the extended phase space. In the phase space, the cosmological constant is connected with the thermodynamic pressure, and the quantity conjugate with pressure in the first law of the thermodynamics is called thermodynamic volume. It means that a term which contains the volume and the variation of the pressure will be factitiously added to the expression of the first law of the black hole thermodynamics. According to the concept of the classical first law of the thermodynamics, the mass of the black hole in the expression can only be considered as the enthalpy, not the internal energy. Considering the hypersurface extends from the event horizon of the AdS black hole to the spatial infinity when the cosmological constant is regarded as a variable, Kastor et al. \cite{Kastor:2009wy} generally construct the Smarr formula and the first law of the AdS black hole thermodynamics by calculating the surface integral of a two-form potential for the static Killing field on the boundary of the hypersurface. The expression of the first law of the AdS black hole thermodynamics in the extended phase space can be strictly derived. Moreover, the thermodynamic volume can be obtained through integrating the Killing potential on the boundary, while its physical significance can naturally understand. The thermodynamic volume is actually that the volume outside the event horizon minus the volume of a spatial slice of the AdS spacetime. Besides, the results also generally indicate that in the extended phase space, the mass of the AdS black hole in the first law should be considered as the enthalpy. Based on it, using the previous method, i.e., a test particle is absorbed by the black hole, the WCCC for some AdS black holes has been examined in the extended phase space \cite{Chen:2019nsr, Zeng:2019huf, Han:2019kjr, Han:2019lfs, Zeng:2019hux, He:2019fti, Hu:2020ccj}. Although the cosmological constant is considered as a variable in these investigations, the particle and the spacetime do not regard as a complete dynamical system as mentioned above. Therefore, in our research, based on the perspective proposed by Sorce and Wald, while the cosmological is regarded as a dynamical variable, we would like to examine the WCCC for RN-AdS black holes under the second-order approximation of the perturbation that comes from the matter fields.

The organization of the paper is as follows. In Sec. \ref{sec2}, we discuss the spacetime geometry of RN-AdS black holes under the matter fields perturbation. In Sec. \ref{sec3}, based on the new version of the gedanken experiment, we derive the first- and the second-order perturbation inequalities for the RN-AdS black hole. In Sec. \ref{sec4}, using the first-order optimal option and the second-order perturbation inequality, we examine the WCCC for nearly extremal RN-AdS black holes under the second-order approximation of the matter fields perturbation. The paper ends with discussions and conclusions in Sec. \ref{sec5}.

\section{Perturbed geometry of RN-AdS black holes}\label{sec2}
Considering the four-dimensional Einstein-Maxwell-AdS gravitational theory, the Lagrangian four-form is given by
\begin{equation}\label{PL}
	\boldsymbol{L}=\frac{1}{16 \pi}\left(R-2 \Lambda-F_{a b} F^{a b}\right) \boldsymbol{\epsilon},
\end{equation}
where $\boldsymbol{F} = d \boldsymbol{A}$ is the electromagnetic strength, $\boldsymbol{A}$ is the vector potential of the electromagnetic field, $R$ is the Ricci scalar, and $\Lambda$ is the cosmological constant with negative value. From the Lagrangian, a class of static spherically symmetric solutions describing RN-AdS black holes is
\begin{equation}
\begin{aligned} \label{metric}
	d s^{2} &=-f(r) d v^{2}+2 d v d r+r^{2}\left(d \theta^{2}+\sin ^{2} \theta d \phi^{2}\right), \\
    \boldsymbol{A} &=-\frac{Q}{r} d v, \qquad \boldsymbol{F} = \frac{Q}{r^2} d r \wedge d v,
\end{aligned}
\end{equation}
where
\begin{equation}\label{fr}
	f(r)=1-\frac{2 M}{r}+\frac{Q^{2}}{r^{2}}-\frac{ \L r^{2}}{3}
\end{equation}
is the blackening factor. The parameters $M$ and $Q$ are associated with the physical mass and electric charge of the black hole. The radius of the event horizon $r_h$ is corresponding to the largest root of the equation $f(r)=0$. Furthermore, the surface gravity, the area of the event horizon, and the electric potential are given as follows:
\ba\begin{aligned}
\kappa=\frac{f^{\prime}\left(r_{h}\right)}{2}, \quad  A_{\mathcal{H}}=4 \pi r_{h}^{2}, \quad \Phi_\mathcal{H}=\frac{Q}{r_{h}}\, .
\end{aligned}\ea

Subsequently, we would like to consider a process that matter fields outside the black hole pass through the event horizon and drop into the black hole. In the process, we propose an assumption that the cosmological constant can be regarded as an effective parameter determined by the matter source coupled to the Einstein-Maxwell gravity. Under this assumption, the matter fields and the spacetime of black holes can be treated as a complete dynamical system, the Lagrangian for the system can be rewritten as
\begin{equation}
	\boldsymbol{L}=\frac{1}{16 \pi}\left(R-F_{a b} F^{a b}\right) \boldsymbol{\epsilon}+\bm{L}_\text{mt},
\end{equation}
where $\bm{L}_\text{mt}$ represents the Lagrangian of the matter fields. The Einstein-Maxwell-AdS gravity can be efficiently derived from above Lagrangian if there is a stable solution of the matter fields such that
\begin{equation}\label{CT1}
T_{ab}=\frac{\Lambda}{8 \pi} g_{ab}\,,
\end{equation}
where $T_{ab}$ is the stress-energy tensor of the matter fields. Since the cosmological constant is contained in the matter fields, its value will change with the matter fields falling into the black hole. In other words, the cosmological constant can be seen as a variable during the process. When the cosmological constant is regarded as a variable, it will relate to the thermodynamic pressure according the black hole thermodynamics in the extended phase space. While based on the first law of thermodynamics in the space, the quantity conjugate with the pressure is defined as a thermodynamic volume. Therefore, utilizing the radius of the event horizon again, the thermodynamic pressure and volume can be expressed as
\begin{equation}
	P=-\frac{\Lambda}{8 \pi}, \qquad V_\mathcal{H} = \frac{4}{3} \pi r_h^3 \,.
\end{equation}

To simplify this discussion, we only consider the configuration of matter fields is spherically symmetric. The configurations of the metric $g_{ab}$, electromagnetic field $\boldsymbol{A}$, and the matter fields can be uniformly labeled as a symbol $\varphi$. The changing behavior of the field configurations with the process can be described by the one-parameter family, i.e., $\varphi(\lambda)$. When $\lambda = 0$, the configuration $\varphi(0)$ represents the background spacetime which is a RN-AdS black hole. If the value of $\lambda$ is small enough, the process can be treated as a perturbation. In this parameter family, the equation of motion can be written as
\begin{equation}
\begin{split}
    &R_{a b}(\lambda)-\frac{1}{2} R(\lambda) g_{a b}(\lambda)=8 \pi\left[T_{a b}^{\mathrm{EM}}(\lambda)+T_{a b}(\lambda)\right], \\
    &\nabla_{a}^{(\lambda)} F^{ba}(\lambda)=4 \pi j^{a}(\lambda).
\end{split}
\end{equation}
where $T_{ab}^\text{EM}$ is the stress-energy tensor of the electromagnetic field as follows:
\ba\begin{aligned}
T_{ab}^\text{EM}&=\frac{1}{4\p}\lf[F_{ac}F_b{}^c-\frac{1}{4}g_{ab}F_{cd}F^{cd}\rt]\,.
\end{aligned}\ea

During the process of the perturbation, the spacetime geometry can be generally described as
\begin{equation}\label{merticdurnper}
	ds^2 = - f(v, r, \lambda) dv^2 + 2 \mu (v, r, \lambda ) d v d r + r^2 (d\theta^2 + \sin^2 \theta d \phi^2),
\end{equation}
where $\mu(v, r, \lambda)$ is a function of arbitrary form. When $f(v, r, 0) = f (r)$ and $\mu (v, r, 0) = 1$, the metric will degenerate into the background spacetime.

Based on the same perspective as Sorce and Wald, we only focus on the case that the perturbation is vanishing on the bifurcation surface $B$ and satisfying a \textbf{stability condition}. The stability condition states that at the sufficiently late times, the spacetime geometry can also be described by the class of RN-AdS solutions, just the parameters $M$, $Q$, and $\Lambda$ are all replaced to $M(\lambda)$, $Q(\lambda)$, and $\Lambda(\lambda)$, i.e.,
\ba
\begin{aligned}\label{S1dsa}
ds^{2}(\lambda) &=-f(r, \lambda) d v^{2}+ 2 d v d r +r^{2}\left(d \theta^{2}+\sin ^{2} \theta d \phi^{2}\right), \\
\boldsymbol{A}(\lambda) &= - \frac{Q(\lambda)}{r} d v, \qquad \boldsymbol{F} (\lambda) = \frac{Q(\lambda)}{r^2} d r \wedge d v,
\end{aligned}\ea
where the blackening factor $f(r, \lambda)$ can be written as
\begin{equation}\label{blackening2}
	f(r, \lambda)=1-\frac{2 M(\lambda)}{r}+\frac{Q^{2}(\lambda)}{r^{2}}-\frac{\L(\l) r^{2}}{3}\,.
\end{equation}
Besides, according to the stability condition, the stress-energy tensor of the matter fields at sufficiently late times should be given as
\begin{equation}
T_{a b}(\l)=\frac{\Lambda(\l)}{8\pi} g_{ab}(\l)\,.
\end{equation}

To ensure above properties of the spacetime and the matter fields, the current which contains the matter fields with the energy, the electric charge, and the cosmological constant must pass through a finite portion of the future event horizon and entry into the black hole. Therefore, we can always choose a hypersurface $\Sigma=\math{H}\cup \S_1$. Here $\mathcal{H}$ is a part of the event horizon which is starting from the bifurcation surface and continuing up the future horizon until a two-dimensional cross section $B_1$ at sufficient late times, and $\Sigma_1$ is starting from $B_1$ and along the time-slice to go to the infinity. In this choice, the hypersurface $\Sigma$ is independent on the parameter $\lambda$. It means that $r_h$ is only the radius of the event horizon of the background spacetime. According to the stability condition, the spacetime geometry on the hypersurface $\Sigma_1$ can be described by Eq. \eq{S1dsa}.

\section{Derivation of the perturbation inequalities}\label{sec3}
Starting from this section, we will examine the WCCC for RN-AdS black holes under the perturbation that comes from the matter fields. Before examination the WCCC, we should derive the first- and the second-order perturbation inequalities. To obtain the two perturbation inequalities, we will consider the off-shell variation of the Einstein-Maxwell gravitational theory based on Iyer-Wald formalism \cite{IW}. We focus on a general diffeomorphism-covariant theory on a four-dimensional spacetime with the Lagrangian four-form $\bm{L}$. In the Einstein-Maxwell gravity, the dynamical fields consist of a Lorentz signature metric $g_{ab}$ and the electromagnetic field $\bm{A}$. As mentioned above, we use the symbol $\phi$ to represent the collection of the dynamical field operators, i.e., $\phi=(g_{ab}, \bm{A})$. When considering the perturbation of the spherically symmetric matter fields, the changing behavior of the configuration of the fields can be described by the one-parameter family, i.e., $\f(\l)$. Moreover, the variation of the quantity $\h$ related to the field $\f$ is defined by
\ba\begin{aligned}
\d\h=\left.\frac{d \h(\l)}{d\l}\right|_{\l=0}\,,\quad\quad
\d^2\h=\left.\frac{d^2 \h(\l)}{d\l^2}\right|_{\l=0}\,.
\end{aligned}\ea
Using the method of the off-shell variation, the variation of $\boldsymbol{L}$ can be generally given by
\begin{equation}\label{varL}
	\delta \boldsymbol{L}=\boldsymbol{E}_{\phi} \delta \phi+d \boldsymbol{\Theta}(\phi, \delta \phi),
\end{equation}
where $\boldsymbol{E}_\phi=0$ is the equation of motion of the on-shell fields related to the Lagrangian $\bm{L}$, and $\boldsymbol{\Theta}$ is called the symplectic potential three-form which is locally constructed out of $\phi$ and its derivatives. For Einstein-Maxwell gravity, the Lagrangian is
\ba\begin{aligned}
\bm{L}=\frac{1}{16 \pi}\left(R-F_{a b} F^{a b}\right)\bm{\epsilon}\,.
\end{aligned}\ea
According to Eq. (\ref{varL}), the expressions of the $\boldsymbol{E}_\phi$ and $\boldsymbol{\Theta}$ are formally given by
\begin{equation}\label{eom}
	\boldsymbol{E}(\phi) \delta \phi=-\boldsymbol{\epsilon}\left[\frac{1}{2} T^{a b} \delta g_{a b}+j^{a} \delta A_{a}\right],
\end{equation}
and
\begin{equation}\label{sypo}
	\begin{split}
 		\boldsymbol{\Theta}_{a b c}^\text{GR}(\phi, \delta \phi)&=\frac{1}{16 \pi} \bm{\epsilon}_{d a b c} g^{d e} g^{f g}\left(\nabla_{g} \delta g_{e f}-\nabla_{e} \delta g_{f g}\right), \\
 		\boldsymbol{\Theta}_{a b c}^\text{EM}(\phi, \delta \phi)&=-\frac{1}{4 \pi} \bm{\epsilon}_{d a b c} F^{d e} \delta A_{e}\,,
	\end{split}
\end{equation}
with
\begin{equation}\label{TJ}
	T_{a b}=\frac{1}{8 \pi}\left(R_{a b}-\frac{1}{2} R g_{a b}\right)-T_{a b}^{\mathrm{EM}}, \quad j^{a}=\frac{1}{4 \pi} \nabla_{a} F^{ba}\,,
\end{equation}
where $T_{ab}$ and $j^a$ can be treated as the stress-energy tensor and the electric current of the matter fields. For the background spacetime, the stress-energy tensor has the form as Eq. \eq{CT1}.

Based on the symplectic potential, the symplectic current three-form is defined as
\begin{equation}
	\boldsymbol{\omega}\left(\phi, \delta_{1} \phi, \delta_{2} \phi\right)=\delta_{1} \boldsymbol{\Theta}\left(\phi, \delta_{2} \phi\right)-\delta_{2} \boldsymbol{\Theta}\left(\phi, \delta_{1} \phi\right)\,.
\end{equation}
Since the symplectic potential can be linearly decomposed as the gravity part and electromagnetic part, the symplectic current can be decomposed the two parts as well, i.e.,
\begin{equation} \boldsymbol{\omega}=\boldsymbol{\omega}^\text{GR}+\boldsymbol{\omega}^\text{EM}.
\end{equation}
The specific expressions of the two parts are respectively given as
\begin{equation}
\begin{split}
  \boldsymbol{\omega}_{a b c}^\text{GR} &=\frac{1}{16 \pi} \bm{\epsilon}_{d a b c} w^{d}, \\
  \bm{\omega}_{a b c}^\text{EM} &=\frac{1}{4 \pi}\left[\delta_{2}\left(\bm{\epsilon}_{d a b c} F^{d e}\right) \delta_{1} A_{e}-\delta_{1}\left(\bm{\epsilon}_{d a b c} F^{d e}\right) \delta_{2} A_{e}\right] ,
\end{split}
\end{equation}
where we have denoted
\ba\begin{aligned}
	w^{a}=P^{a b c d e f}\left(\delta_{2} g_{b c} \nabla_{d} \delta_{1} g_{e f}-\delta_{1} g_{b c} \nabla_{d} \delta_{2} g_{e f}\right)
\end{aligned}\ea
with
\ba\begin{aligned}
	P^{a b c d e f} &=g^{a e} g^{f b} g^{c d}-\frac{1}{2} g^{a d} g^{b e} g^{f c}-\frac{1}{2} g^{a b} g^{c d} g^{e f} \\
&-\frac{1}{2} g^{b c} g^{a e} g^{f d}+\frac{1}{2} g^{b c} g^{a d} g^{e f}.
\end{aligned}\ea
Let $\zeta^a$ be a generator of the deffeomorphism. Replacing $\delta$ with $\mathcal{L}_\zeta$ in Eq. \eq{varL}, one can define the Noether current three-form $\boldsymbol{J}_\zeta$ associated with $\zeta^a$,
\begin{equation}\label{Jz1}
	\boldsymbol{J}_{\zeta}=\boldsymbol{\Theta}\left(\phi, \mathcal{L}_{\zeta} \boldsymbol{\phi}\right)-\zeta \cdot \boldsymbol{L}.
\end{equation}
On the other hand, it was shown in Ref. \cite{Wald94} that the Noether current can also be formally written as
\begin{equation}\label{Jz2}
	\boldsymbol{J}_{\zeta}=\boldsymbol{C}_{\zeta}+d \boldsymbol{Q}_{\zeta},
\end{equation}
where $\boldsymbol{Q}_\zeta$ is called the Noether charge and $\boldsymbol{C}_\zeta = \zeta^a \boldsymbol{C}_a$ are the constraints of the theory. When $\boldsymbol{C}_a = 0$, the dynamical fields will satisfy the equation of motion. In the theory of Einstein-Maxwell gravity, the Noether charge can also be decomposed into two parts which represent the conservation charge of the gravity and the electromagnetic respectively, i.e.,
\begin{equation}\label{tnc}
	\bm{Q}_\zeta = \bm{Q}_\zeta^\text{GR}+\boldsymbol{Q}_\zeta^\text{EM},
\end{equation}
where the specific expression of the $\boldsymbol{Q}_\zeta^\text{GR}$ and $\boldsymbol{Q}_\zeta^\text{EM}$ can be given as
\begin{equation}\label{QQQ}
\begin{split}
	\left(\bm{Q}_{\zeta}^{\mathrm{GR}}\right)_{a b} &=-\frac{1}{16 \pi} \bm{\epsilon}_{a b c d} \nabla^{c} \zeta^{d}, \\
	\left(\bm{Q}_{\zeta}^{\mathrm{EM}}\right)_{a b} &=-\frac{1}{8 \pi} \bm{\epsilon}_{a b c d} F^{c d} A_{e} \zeta^{e},
\end{split}
\end{equation}
and the constraints are defined as
\begin{equation}\label{constraints}
	\bm{C}_{a b c d}=\bm{\epsilon}_{e b c d}\left(T_{a}{}^{e}+A_{a} j^{e}\right).
\end{equation}

Due to the perturbation that comes from the spherically symmetric matter fields, we consider the diffeomorphism generated by the static Killing vector field $\x^a=(\pd/\pd v)^a$ on the background geometry. By virtue of the diffeomorphism, we can choose a gauge such that the coordinates $(v, r, \q, \f)$ are fixed under the variation. It means that the Killing vector is invariable during the process of the perturbation, i.e., $\delta \xi^a = 0$. Taking the first-order variations on Eq. \eq{Jz1} and Eq. \eq{Jz2}, one can obtain the first-order variational identity,
\begin{equation}\label{fvi}
	d\left[\delta \boldsymbol{Q}_{\x}-\x \cdot \boldsymbol{\Theta}(\phi, \delta \phi)\right]=\bm{\omega}\left(\phi, \delta \phi, \mathcal{L}_{\x} \phi\right)-\x \cdot \boldsymbol{E}_{\phi} \delta \phi-\delta \bm{C}_{\x}.
\end{equation}
After that, taking the variation on above equality again and using the fact that $\mathcal{L}_\x \phi =0$ for the background fields, the second-order variational identity can be also obtained as
\ba\begin{aligned}\label{var2}
	&d\left[\delta^{2} \boldsymbol{Q}_{\x}-\x \cdot \delta \boldsymbol{\Theta}(\phi, \delta \phi)\right]\\ &=\boldsymbol{\omega}\left(\phi, \delta \phi, \mathcal{L}_{\x} \delta \phi\right)-\x \cdot \boldsymbol{E}_{\phi} \delta^{2} \phi-\x \cdot \delta \boldsymbol{E}_{\phi} \delta \phi-\delta^{2} \boldsymbol{C}_{\x}\,.
\end{aligned}\ea

Next, we will calculate the integral from of the first- and the second-order variational identity respectively. For the first-order variational identity, integrating it on the hypersurface $\Sigma$ and utilizing the condition $\math{L}_\x\f=0$, we have
\ba\begin{aligned}\label{eq1}
\int_{\S}d\lf[\d \bm{Q}_\x-\x\.\bm{\Q}(\f,\d\f)\rt]+\int_{\S}\x\.\bm{E}_\f\d\f+\int_{\S}\d\bm{C}_\x=0\,.
\end{aligned}\nn\\\ea
Following the similar consideration in Ref. \cite{Sorce:2017dst}, we impose the gauge condition of the electromagnetic field such that $\x^a\d A_a=0$ on the event horizon $\math{H}$. According to the calculation in Appendix A, the integral form of the first-order variational identity can be obtained as
\ba\begin{aligned}\label{var1ineq}
\d M-\F_\mathcal{H} \d Q - V_\mathcal{H} \delta P = \delta \left[\int_{\mathcal{H}} \tilde{\boldsymbol{\epsilon}} \mu (v, r_h) T_{ab} \xi^a (dr)^b \right] \,.
\end{aligned}\ea
To obtain the first-order perturbation inequality, we should find out the relation between the right-hand side of Eq. (\ref{var1ineq}) and the null energy of the matter fields. Since we consider the perturbation that comes from the spherically symmetric matter fields, a null vector field during the perturbation can be chosen as
\begin{equation}
	l^a (\lambda) = \xi^a + \beta (\lambda) r^a,
\end{equation}
where
\begin{equation}
	r^a = \left(\frac{\partial}{\partial r} \right)^a, \qquad \beta (\lambda) = \frac{f (v, r_h, \lambda)}{2 \mu (v, r_h, \lambda)}.
\end{equation}
The null energy condition under the process can be expressed as
\begin{equation}
	T_{ab}(\lambda)  l^a (\lambda) l^b (\lambda) \ge 0.
\end{equation}
It can be demonstrated that the null energy condition satisfied the following relation
\begin{equation}\label{nullenergycondmubeta}
	\begin{split}
		& T_{ab}(\lambda)  l^a (\lambda) l^b (\lambda) \\
		& = \mu (v, r_h, \lambda) T_{ab} (\lambda) \xi^a (dr)^b + \beta (\lambda)^2 T_{ab} (\lambda) r^a r^b.
	\end{split}
\end{equation}
Considering the fact $\beta(0) = 0$ for the background spacetime, the null energy condition under the first-order approximation of the perturbation can be obtained as
\ba\label{fonullencon}\begin{aligned}
&\int_\mathcal{H} T_{ab} (\lambda) l^a (\lambda) l^b (\lambda) dv \wedge \tilde{\boldsymbol{\epsilon}} \\
&\simeq \lambda \delta \left[\int_{\mathcal{H}} \tilde{\boldsymbol{\epsilon}} \mu (v, r_h) T_{ab} \xi^a (dr)^b \right] \ge 0.
\end{aligned}\ea
Substituting Eq. (\ref{fonullencon}) into Eq. (\ref{var1ineq}), the integral form of the first-order variational identity can be reduced as
\begin{equation}
	\delta M - \Phi_\mathcal{H} \delta Q - V_\mathcal{H} \delta P \ge 0.
\end{equation}
This inequality is called the first-order perturbation inequality.

In our investigation, we wish to examine the WCCC for RN-AdS black holes under the second-order approximation of the matter fields perturbation. When the first-order inequality is satisfied, we can demonstrate the WCCC cannot be violate under the first-order approximation, while the higher-order approximation can be largely neglected. However, if the first-order perturbation inequality is chosen as an optimal option, i.e.,
\begin{equation}
	\delta M - \Phi_\mathcal{H} \delta Q - V_\mathcal{H} \delta P = 0,
\end{equation}
the WCCC will not be examined only considering the first-order approximation. In this situation, the second-order approximation of the perturbation should be involved. Besides, this optimal condition also implies that the energy flux through the event horizon vanishes under the first-order approximation.

To consider the second-order approximation of the perturbation, we should derive the second-order perturbation inequality. Analogous to derive the first-order perturbation inequality, we can obtain the integral form of the second-order variational identity. Integrating Eq. \eq{var2} on the hypersurface $\S$, we have
\ba\begin{aligned}\label{secvar}
	\int_{\S}d&\left[\delta^{2} \boldsymbol{Q}_{\x}-\x \cdot \delta \boldsymbol{\Theta}(\phi, \delta \phi)\right]=\math{E}_{\S_1}(\f,\d\f)+\math{E}_\math{H}(\f,\d\f)\\ &\quad\quad\quad\quad-\int_{\S}\d\left(\x\.\bm{E}_\f\d\f\right)-\int_{\S}\d^2\bm{C}_\x\,,
\end{aligned}\ea
where we denote
\begin{equation}
	\begin{split}
		\boldsymbol{\mathcal{E}}_{\Sigma_1}(\phi, \delta \phi)=\int_{\Sigma_1}\boldsymbol{\omega}\left(\phi, \delta \phi, \mathcal{L}_{\xi} \delta \phi\right), \\
		\boldsymbol{\mathcal{E}}_{\mathcal{H}}(\phi, \delta \phi)=\int_{\math{H}}\boldsymbol{\omega}\left(\phi, \delta \phi, \mathcal{L}_{\xi} \delta \phi\right).
	\end{split}
\end{equation}
According to the calculation in Appendix B, the second-order variational identity can reduce to
\ba\begin{aligned}\label{Var2}
\d^2M&-\F_\mathcal{H} \d^2Q-V_\mathcal{H}\d^2P=\math{E}_{\S_1}(\f,\d\f)+\math{Y}(\f,\d\f)\\
&+ \delta^2 \left[ \int_\mathcal{H} \tilde{\boldsymbol{\epsilon}} \mu (v, r_h) \left(T_{ab}^{\text{EM}} + T_{ab}\right) (dr)^a \xi^b \right]\,,
\end{aligned}\ea
where we have defined
\ba\begin{aligned}
\math{Y}(\f,\d\f)=\frac{1}{8\p}\int_{S_c}\bm{\epsilon}_{abcd}\d F^{cd}\d A_e\x^e\,.
\end{aligned}\ea

Therefore, the calculation of the integral form of the second-order variational identity only remains to evaluate $\math{E}_{\S_1}(\f,\d\f)$ and $\math{Y}(\f,\d\f)$. According to the calculation in Appendix C, we can obtain the sum of two terms as
\ba\begin{aligned}\label{resultey}
\math{E}_{\S_1}(\f,\d\f)+\math{Y}(\f,\d\f)=\frac{\delta Q^2}{r_h}\,.
\end{aligned}\ea
Substituting Eq. (\ref{resultey}) into Eq. (\ref{Var2}), the integral form of the second-order variational identity can be written as
\begin{equation}\label{secondvariden}
	\begin{split}
&\delta^2 M - \Phi_\mathcal{H} \delta^2 Q -V_\mathcal{H} \delta^2 P  - \frac{\delta Q^2}{r_h}\\
&= \delta^2 \left[ \int_\mathcal{H} \tilde{\boldsymbol{\epsilon}} \mu (v, r_h) \left(T_{ab}^{\text{EM}} + T_{ab}\right) (dr)^a \xi^b \right].
	\end{split}
\end{equation}
According to Eq. (\ref{nullenergycondmubeta}) and $T_{ab}(0) = 0$ for the background spacetime, while using the optimal option of the first-order approximation, the null energy condition under the second-order approximation can be expanded as
\begin{equation}\label{nullenergysecond}\begin{split}
&\int_\mathcal{H} T_{ab}(\lambda) l^a (\lambda) l^b (\lambda) dv \wedge \tilde{\boldsymbol{\epsilon}} \\
&\simeq \frac{\lambda^2}{2} \delta^2 \left(\int_\mathcal{H} \tilde{\boldsymbol{\epsilon}} \mu (v, r_h) T_{ab} (dr)^a \xi^b \right) \ge 0.
\end{split}\end{equation}
Finally, using the result of Eq. (\ref{nullenergysecond}), Eq. (\ref{secondvariden}) can be reduced as
\begin{equation}
	\begin{split}
		\delta^2 M - \Phi_\mathcal{H} \delta^2 Q -V_\mathcal{H} \delta^2 P  - \frac{\delta Q^2}{r_h} \ge 0.
	\end{split}
\end{equation}
It is called the second-order perturbation inequality.

\section{Gedanken experiment to examine the WCCC for a nearly extremal RN-AdS black hole}\label{sec4}
In this section, we will perform two perturbation inequalities to examine the WCCC for nearly extremal RN-AdS black holes under the second-order approximation of the perturbation. Due to the stability condition, the spacetime geometry on the hypersurface $\Sigma_1$ can still belong to the class of RN-AdS solutions. If the geometry on $\Sigma_1$ can be described by the metric of RN-AdS black holes, it will demonstrate that the WCCC cannot be violated after the perturbation. This issue is equivalent to check whether the event horizon exists at the end of the perturbation.
\begin{figure*}
	\begin{center}
	\includegraphics[width=0.49\textwidth,height=0.28\textheight]{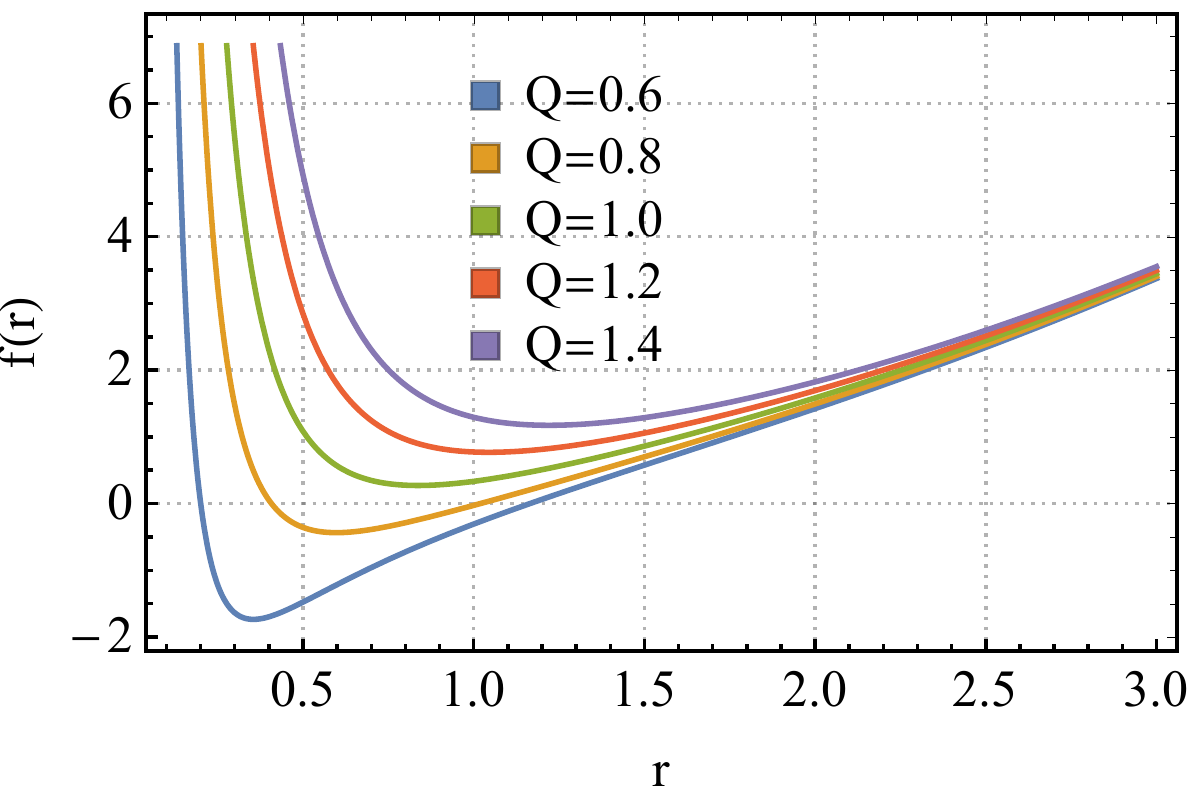}
	\includegraphics[width=0.49\textwidth,height=0.28\textheight]{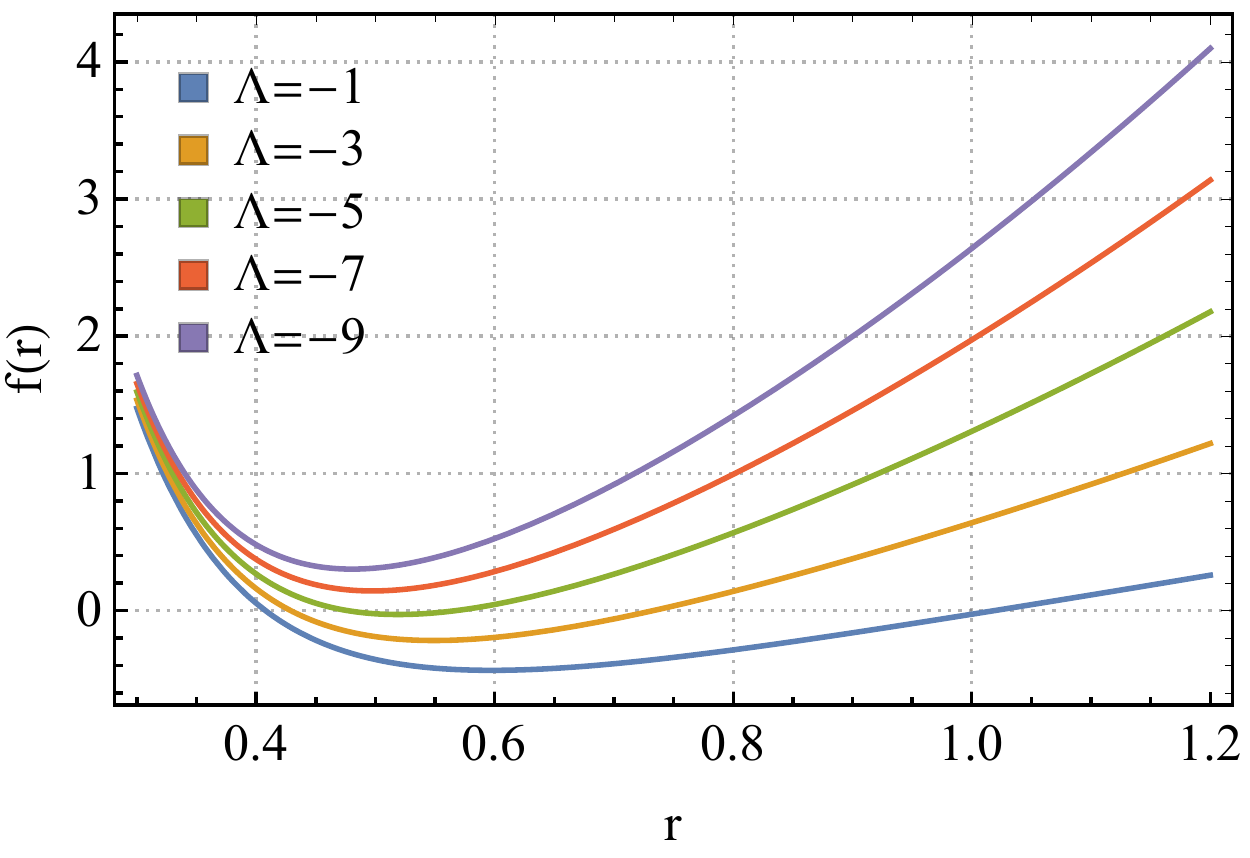}
	\caption{Plot showing the radial dependence of blackening factor $f(r)$ for different values of electric charge $Q$ (left panel) and cosmological constant $\L$ (right panel). In the left panel, we have set $M=1$ and $\L=-1$. In the right panel, we have set $M=1$ and $Q=0.8$}\label{fig}
	\end{center}
\end{figure*}
From \fig{fig}, it is shown that the event horizon exists only if the minimal value of the function $f(r)$ is negative. When the minimal value becomes positive, the event horizon will disappear and the naked singularity will be exposed to spacetime. It means that examining the WCCC for RN-AdS black holes is equivalent to check the sign of the minimal value of the blackening factor $f(r,\l)$ in Eq. \eq{blackening2}. Therefore, we define a function
\ba\label{cd1}
h(\l)=f\lf(r_m(\l),\l\rt)
\ea
to represent the minimal value of blackening factor after the perturbation. In which, $r_m(\l)$ is defined as the position of the minimal value of $f(r, \l)$, and it can be determined by
\ba\begin{aligned}\label{rmeq}
\pd_rf\lf(r_m(\l),\l\rt)=0\,.
\end{aligned}\ea
From Eq. (\ref{rmeq}), the relation between the mass $M$ and the position of the minimal value $r_m$ can be given as
\ba\begin{aligned}
M=\frac{3Q^2+r_m^4\L}{3r_m}\,.
\end{aligned}\ea
Considering the second-order approximation of the perturbation, we can expand the function $h(\lambda)$ with respect to the parameter $\lambda$ at $\lambda = 0$. The specific expression of $h(\lambda)$ under the second-order approximation is
\ba\begin{aligned}
&h(\l)=\frac{r_m^2-Q^2-r_m^4\L}{r_m^2}-\frac{2 \l}{r_m}\lf(\d M-\frac{Q}{r_m}\d Q+\frac{r_m^3}{6}\d \L\rt)\\
&-\frac{\l^2}{r_m}\lf[\d^2M-\frac{Q}{r_m}\d^2 Q+\frac{r_m^3}{6}\d^2 \L+\left(\frac{Q^2}{r_m^3}-r_m\L\right)\d r_m^2\rt]\\
&\l^2\lf(\frac{\d Q^2+2\d M \d r_m}{r_m^2}-\frac{2r_m \d r_m \d \L}{3}-\frac{4Q\d Q\d r_m}{r_m^3}\rt)\,.
\end{aligned}\ea
Taking the first-order variation on Eq. \eq{rmeq}, we can get
\ba\begin{aligned}
\d r_m=\frac{r_m^2}{r_m^2\L-Q^2}\lf(\d M-\frac{2q\d Q}{r_m}-\frac{r_m^3\d \L}{3}\rt)\,.
\end{aligned}\ea

Following a similar setup with Ref. \cite{Sorce:2017dst}, we consider the case that the radius of the event horizon $r_h$ and the position of the minimal value $r_m$ satisfy a reliation $r_m=(1-\epsilon)r_h$ where the parameter $\epsilon$ is small enough which is chosen to agree with the first-order approximation of the matter fields perturbation. Combining the relation with Eq. \eq{rmeq}, one can obtain
\ba
f'\lf(r_h\rt)=\epsilon r_h f''\lf(r_h\rt)
\ea
under the first-order approximation of $\epsilon$. Meanwhile, the minimal value of $f(r)$ under the second-order approximation of $\epsilon$ can also be expressed as
\ba\begin{aligned}
&f(r_m)=f\lf(r_h(1-\epsilon)\rt)\\
&\simeq -\epsilon r_h f'(r_h)+\frac{\epsilon^2 r_h^2}{2} f''(r_h)\\
&=-\frac{1}{2}\epsilon^2 r_h^2 f''(r_h)\,,
\end{aligned}\ea
which gives the following expression
\ba\begin{aligned}
\frac{r_m^2-q^2-r_m^4\L}{r_m^2}=\lf(1-\frac{2Q^2}{r_h^2}\rt)\epsilon^2\, .
\end{aligned}\ea

Utilizing the optimal option of the first-order approximation and the second-order perturbation inequality, while combining above results, the minimal value of $f(r,\l)$ under the second-order approximation can reduce to
\ba\begin{aligned}\label{hl}
h(\l)&\leq\lf(1-\frac{2Q^2}{r_h^2}\rt)\epsilon^2+\frac{\l \epsilon}{r_h^2}\lf(2Q\d Q+r_h^4\d\L\rt)\\
&+\frac{\l^2\lf(2Q\d Q+r_h^4\d \L\rt)^2}{4r_h^2(\L r^6_h-Q^2)}\,.
\end{aligned}\ea
Solving the equations $f((1+\epsilon)r_m)=0$ and $f'(r_m)=0$ under the zero-order approximation of $\epsilon$, we have
\ba\begin{aligned}\label{extremal}
\L=\frac{r_m^2-Q^2}{r_m^4}\,,\ \ M=\frac{2Q^2+r_m^2}{3r_m}\,.
\end{aligned}\ea
Substituting Eq. (\ref{extremal}) into Eq. \eq{hl}, the expression of $h(\lambda)$ under the second-order approximation can be rewritten as
\ba\begin{aligned}\label{hl2}
h(\l)&\leq-\frac{(2Q\d Q \lambda+r_m^4\d \L \lambda - 4Q^2\epsilon+2r_m^2\epsilon)^2}{2r_m^6f''(r_m)}\,.
\end{aligned}\ea
Since $r_m$ is the minimal value of $f(r)$, we have $f''(r_m)>0$. Therefore, Eq. \eq{hl2} means that the value of $h(\lambda)$ is negative, i.e., $h(\lambda) \leq 0$. This result implies that the event horizon of RN-AdS black holes still exists after the perturbation, and the WCCC cannot be violated under the second-order approximation of the matter fields perturbation.

\section{Conclusions}\label{sec5}
Based on the new version of the gedanken experiment proposed by Sorce and Wald, we examine the validity of the WCCC for RN-AdS black holes under the second-order approximation of the perturbation that comes from the spherically symmetric matter fields. The matter fields are considered to be sufficiently coupled to the black hole spacetime, and the cosmological constant is regarded as a portion of the matter fields. It implies that the cosmological constant can be seen as a dynamical variable, its value can vary with the matter fields perturbing to the black hole. To examine the WCCC after the perturbation, we propose the stability condition which states that the spacetime geometry still belongs to the class of static spherically symmetric solutions describing RN-AdS solutions at sufficiently time. Based on the stability condition and the null energy condition, the first two order perturbation inequalities are obtained. In the expressions of the two inequalities, since the cosmological constant is treated as a variable, the inequalities are first extended to the case containing the thermodynamic pressure and volume. Utilizing the first-order optimal option and the second-order perturbation inequality, together with the stability condition and the null energy condition, the WCCC for nearly extremal RN-AdS black holes after the second-order approximation of the perturbation is examined. It is shown that under the second-order approximation, the event horizon of the RN-AdS black hole still exits, and the naked singularity does not appear in the spacetime. This result sufficiently illustrates that the WCCC cannot be violated under the second-order approximation of the matter fields perturbation.

\section*{Acknowledgement}
This research was supported by National Natural Science Foundation of China (NSFC) with Grants No. 11775022 and 11873044..

\section*{Appendix A: Derivation of the first order variational identity}\label{secA}

In this Appendix, we would like to calculate the integral form of the first-order variational identity on the hypersurface $\Sigma$ in Eq. \eq{eq1}. Using the property $\Sigma = \mathcal{H} \cup \Sigma_1$ and the Stoke's theorem, it can be written as
\ba\begin{aligned}\label{varA}
&\int_{S_c}\d \left[\bm{Q}_\x-\x\.\bm{\Theta}(\f,\d\f)\right]-\int_B\d \left[\bm{Q}_\x-\x\.\bm{\Theta}(\f,\d\f)\right]\\
&+\int_{\S_1}\x\.\bm{E}_\f\d\f+\int_{\S_1}\d\bm{C}_\x+\int_\math{H}\d\bm{C}_\x=0\,.
\end{aligned}\ea
After regarding the cosmological constant as a dynamical quantity, the divergence will appear in the result of the integration at the asymptotic infinity. Therefore, we choose a cut-off sphere $S_c$ with radius $r_c$ to replace the boundary of the asymptotic infinity of $\S_1$. In the finial results, we will take the limitation such that $S_c$ approaches asymptotic infinity again. According to the assumption introduced in Section \ref{sec2}, we propose that the perturbation is vanishing on the bifurcation surface $B$. Therefore, the second term in Eq. (\ref{varA}) can be directly neglected. Following a similar setup in Ref. \cite{Sorce:2017dst}, we also impose the gauge condition of the electromagnetic field such that $\x^a\d A_a=0$ on $\math{H}$, which implies that we cannot directly perform
\ba\begin{aligned}
\tilde{\bm{A}}(\l)=-\frac{Q(\l)}{r} dv
\end{aligned}\ea
to calculate some quantities related to $\d \bm{A}$ and $\d^2\bm{A}$ on the hypersurface $\S_1$. However, since $\bm{F}(\l)$ is gauge invariant, it is reasonable for us to calculate the quantities on $\S_1$ using the explicit expressions of the electromagnetic strength in Eq. \eq{S1dsa}. The volume element of the spacetime on $\S_1$ can be written as
\ba\begin{aligned}
\bm{\epsilon}(\l)=r^2\sin\q dt\wedge dr\wedge d\q\wedge d\f\,,
\end{aligned}\ea
which implies that $\d \bm{\epsilon}=0$. From Eq. \eq{eom}, the stress-energy and electric current of the matter fields on $\S_1$ can be expressed as
\ba\begin{aligned}\label{TJl}
T_{ab}(\l)=\frac{\L(\l)}{8\p} g_{ab}(\l)\,,\ \
j^a(\l)=0\,.
\end{aligned}\ea

Firstly, we turn to calculate the integrations on the hypersurface $\S_1$. For the third term of Eq. \eq{varA}, substituting Eq. \eq{eom} into it, we have
\ba\begin{aligned}
\int_{\S_1}\x\.\bm{E}_\f\d\f&=-\int_{\S_1}\x\.\bm{\epsilon}\left[\frac{1}{2}T^{ab}\d g_{ab}+j^a\d A_a\right]\\
&=\frac{\L}{16\p}\int_{\S_1}\x\.\bm{\epsilon}g^{ab}\d g_{ab}\,,
\end{aligned}\ea
where we have used Eq. \eq{TJl}. Based on the specific expression of the metric in Eq. \eq{S1dsa}, we can prove the following relation
\begin{equation}\label{gdeltag}
	g^{ab} (\lambda) \frac{dg_{ab} (\lambda)}{d\lambda} = 0\,.
\end{equation}
It means that the integral of the third term is equal to zero.

For the fourth term of Eq. \eq{varA}, we first derive the specific expression of the constraints on $\S_1$. According to Eq. \eq{constraints}, we can obtain
\ba\begin{aligned}\label{Cx}
[\bm{C}_\x(\l)]_{bcd}&=\bm{\epsilon}_{ebcd}\left[T_a{}^e(\l)\x^a+A_a(\l) j^e(\l)\x^a\right]\\
&=\frac{\L(\l)r^2\sin\q}{8\p} (dr)_b\wedge (d\q)_c\wedge (d\f)_d\,.
\end{aligned}\ea
The fourth term can further be obtained as
\ba\begin{aligned}
\int_{\S_1}\d\bm{C}_\x&=\int_{r_h}^{r_c} \frac{r^2\d \L}{2}dr=(V_c-V_H)\d P\,,
\end{aligned}\ea
where we have denoted
\ba\begin{aligned}
V_c=\frac{4}{3}\p r_c^3\,.
\end{aligned}\ea

For the fifth term of Eq. \eq{varA}, using $A_a \xi^a |_\mathcal{H} = - \Phi_\mathcal{H}$, we have
\begin{equation}\label{lastterm}
	\begin{split}
		&\int_\mathcal{H} \delta \boldsymbol{C}_\xi = \delta \int_\mathcal{H} \left[\boldsymbol{\epsilon}_{ebcd} \left(T_a^{\; e} \xi^a + A_a \xi^a j^e \right) \right] \\
		& = - \delta \left(\int_\mathcal{H} \tilde{\boldsymbol{\epsilon}} \mu \left(v, r_h \right) T_{ae} \xi^a (dr)^e \right) - \Phi_\mathcal{H} \delta \left(\int_\mathcal{H} \boldsymbol{\epsilon}_{ebcd} j^e \right)\,,
	\end{split}
\end{equation}
where we have denoted the volume element of the event horizon $\mathcal{H}$ as $\tilde{\boldsymbol{\epsilon}} = dv \wedge \hat{\boldsymbol{\epsilon}}$, and $\hat{\boldsymbol{\epsilon}} = r^2 \sin \theta d \theta \wedge d \phi$ is the volume element of the section of the event horizon. Since the perturbation satisfies the stability condition, all charges contained in the matter fields should pass through $\math{H}$ and fall into the black hole. From this perspective, according to the electromagnetic current of the matter fields in \eq{TJ}, we can further obtain
\begin{equation}
	\begin{split}
		\delta \left(\int_{\mathcal{H}} \boldsymbol{\epsilon}_{ebcd} j^e \right) = & \frac{1}{4 \pi} \delta \left(\int_\mathcal{H} \boldsymbol{\epsilon}_{ebcd} \nabla_a F^{ea} \right) \\
		= & \frac{1}{8 \pi} \delta \left(\int_{B_1} \boldsymbol{\epsilon}_{ebcd} F^{eb}-\int_{B} \boldsymbol{\epsilon}_{ebcd} F^{eb} \right) \\
		= & \delta Q.
	\end{split}
\end{equation}
Therefore, Eq. \eq{lastterm} can be reduces to
\ba\begin{aligned}\label{lastterm2}
\int_\math{H}\d\bm{C}_\x&=- \delta \left(\int_\mathcal{H} \tilde{\boldsymbol{\epsilon}} \mu \left(v, r_h \right) T_{ae} \xi^a (dr)^e \right) - \F_H \d Q\,.
\end{aligned}\ea

Finally, we consider the first term of Eq. \eq{varA}. This term can be linearly decomposed into the gravitational part and electromagnetic part because the Noether charge and the symplectic potential are both linearly decomposed as the two parts. For the gravitational part, combining Eq. \eq{sypo} with Eq. \eq{QQQ}, while using the explicit expression of the metric in Eq. \eq{S1dsa}, we can obtain
\ba\begin{aligned}\label{gravtypart}
&\frac{d\bm{Q}_\xi^\text{GR}(\l)}{d\l} - \xi \cdot \bm{\Theta}^\text{GR} \left(\phi(\l), \frac{d\phi(\l)}{d\l}\right)\\
&=\frac{1}{4\p}\left(M'(\l)+\frac{r^3 \L'(\l)}{6}-\frac{Q(\l)Q'(\l)}{r}\right)\sin\q d\q\wedge d\f\,.
\end{aligned}\ea
Integrating it on the cut-off sphere $S_c$, we have
\begin{equation}\label{lhsi}
\begin{split}
&\int_{S_c} \left[\delta \bm{Q}_\xi^\text{GR} - \xi \cdot \bm{\Theta}^\text{GR} (\phi, \delta \phi) \right] = \delta M -V_c \delta P\,,
\end{split}
\end{equation}
For the electromagnetic part, as mentioned above, we have not the specific expression of $\bm{A}(\l)$ to evaluate $\d \bm{A}$ and $\d^2\bm{A}$ on $\Sigma_1$. Therefore, we only perform the specific expressions of the electromagnetic strength in Eq. \eq{S1dsa} to evaluate this part. According to the electromagnetic part in Eq. \eq{sypo} and Eq. \eq{QQQ}, we can derive the integrated as
\ba\begin{aligned}\label{eq33}
&\left[\frac{d\bm{Q}_\xi^\text{EM}(\l)}{d\l} - \xi \cdot \bm{\Theta}^\text{EM} \left(\phi(\l), \frac{d\phi(\l)}{d\l}\right)\right]_{ab} \\
&=-\frac{1}{8\p}\bm{\epsilon}_{abcd}\left(\frac{ d F^{cd}(\l)}{d\l}A_e(\l)\x^e+F^{cd}(\l)\frac{d A_e(\l)}{d\l}\x^e\right.\\
&\quad\quad\quad\quad\left.-2F^{ae}(\l)\x^b\frac{ A_e(\l)}{d\l}\right)\,,
\end{aligned}\ea
on the hyersurface $\S_1$. Utilizing the expression of the electromagnetic strength in Eq. (\ref{S1dsa}), we can further obtain the following relation
\ba\begin{aligned}
\bm{\epsilon}_{abcd}F^{cd}(\l)&\frac{d A_e(\l)}{d\l}\x^e=2\bm{\epsilon}_{abcd}F^{ae}(\l)\x^b\frac{d A_e(\l)}{d\l}\\
&=2Q(\l)\x^e\frac{d A_e(\l)}{d\l} \sin \q (d\q)_a\wedge (d\f)_b\,.
\end{aligned}\ea
Based on the relation, Eq. \eq{eq33} can be reduced to
\ba\begin{aligned}\label{eq44}
&\left[\frac{d\bm{Q}_\xi^\text{EM}(\l)}{d\l} - \xi \cdot \bm{\Theta}^\text{EM} \left(\phi(\l), \frac{d\phi(\l)}{d\l}\right)\right]_{ab} \\
&=-\frac{1}{8\p}\bm{\epsilon}_{abcd}\frac{ d F^{cd}(\l)}{d\l}A_e(\l)\x^e\,.
\end{aligned}\ea
Therefore, the integral of the electromagnetic part of the first term can be further calculated
\ba\begin{aligned}\label{eq331}
&\int_{S_c} \left[\delta \bm{Q}_\xi^\text{EM} - \xi \cdot \bm{\Theta}^\text{EM} (\phi, \delta \phi) \right]\\ &=-\frac{1}{8\p}\int_{S_c}\bm{\epsilon}_{abcd}\d F^{cd}A_e\x^e\\
&=\frac{1}{4\p}\int_{S_c}\frac{Q\d Q}{r}\sin\q d\q\wedge d\f\\
&=0.
\end{aligned}\ea

Summing above results, the first-order variational identity can be expressed as
\ba\begin{aligned}\label{varidenwonull}
\d M-\F_H \d Q-V_H\d P= \delta \left(\int_\mathcal{H} \tilde{\boldsymbol{\epsilon}} \mu \left(v, r_h \right) T_{ae} \xi^a (dr)^e \right)\,.
\end{aligned}\ea

\section*{Appendix B: Derivation of the second order variational identity}
In this Appendix, we would like to calculate the second-order variational identity in Eq. \eq{secvar}. Using the Stoke's theorem and the property of the hypersurface $\Sigma$, the integral form of the second-order variational identity can be decomposed as
\ba\begin{aligned}\label{varB}
&\int_{S_c}\d \left[\d \bm{Q}_\x-\x\.\bm{\Q}(\f,\d\f)\right]-\int_B\d \left[\d \bm{Q}_\x-\x\.\bm{\Q}(\f,\d\f)\right]\\
&+ \int_{\math{H}}\d^2\bm{C}_\x +\int_{\S_1}\d^2\bm{C}_\x + \int_{\S_1}\d\left(\x\.\bm{E}_\f\d\f\right)\\
&+ \int_{\math{H}}\d\left(\x\.\bm{E}_\f\d\f\right) -\math{E}_\math{H}(\f,\d\f) -\math{E}_{\S_1}(\f,\d\f) =0\,.
\end{aligned}\ea
Since the perturbation that comes from the matter fields is vanishing on $B$, the second term in Eq. \eq{varB} can also be neglected. With similar calculation as Appendix A, using the specific expression of the metric in Eq. \eq{S1dsa}, one can derive the integrated of the gravitational part of the first term in Eq. (\ref{varB}) as
\ba\begin{aligned}\label{GRvar2}
&\d \left[\d \bm{Q}_\x^\text{GR}-\x\.\bm{\Q}^\text{GR}(\f,\d\f)\right]\\
&=\frac{d}{d\l}\left[\frac{d\bm{Q}_\x^\text{GR}(\l)}{d\l}-\x\.\bm{\Q}^\text{GR}\left(\f(\l),\frac{d\f(\l)}{d\l}\right)\right]_{\l=0}\\
&=\frac{1}{4\p}\left(\d^2 M+\frac{r^3 \d^2\L}{6}-\frac{\d Q^2}{r}-\frac{Q\d^2Q}{r} \right) \sin \theta d\theta\wedge d\phi\,.
\end{aligned}\ea
Integrating it on the cut-off sphere $S_c$, we can obtain
\ba\begin{aligned}\label{ftintegratedgr}
\int_{S_c}\d \left[\d\bm{Q}_\x^\text{GR}-\x\.\bm{\Q}^\text{GR}(\f,\d\f)\right]=\d^2M-V_c\d^2P\,.
\end{aligned}\ea
For the electromagnetic part of the first term, according to \eq{eq44}, we have
\ba\begin{aligned}\label{EMvar2}
&\d \left[\d \bm{Q}_\x^\text{EM}-\x\.\bm{\Q}^\text{EM}(\f,\d\f)\right]\\
&=\frac{d}{d\l}\left[\frac{d\bm{Q}_\x^\text{EM}(\l)}{d\l}-\x\.\bm{\Q}^\text{EM}\left(\f(\l),\frac{d\f(\l)}{d\l}\right)\right]_{\l=0}\\
&=-\frac{1}{8\p}\bm{\epsilon}_{abcd}(\d^2 F^{cd}A_e\x^e+\d F^{cd}\d A_e\x^e)\\
&=\frac{1}{4\p}\frac{Q\d^2Q}{r}\sin\q (d\q)_a\wedge(d\f)_b-\frac{1}{8\p}\bm{\epsilon}_{abcd}\d F^{cd}\d A_e\x^e \,.
\end{aligned}\ea
Therefore, the integral of the electromagnetic part of the first term becomes
\ba\begin{aligned}\label{ftintegratedem}
\int_{S_c}\d \left[\d\bm{Q}_\x^\text{EM}-\x\.\bm{\Q}^\text{EM}(\f,\d\f)\right]=-\math{Y}(\f,\d \f)\,,
\end{aligned}\ea
where we have denoted
\ba\begin{aligned}
\math{Y}(\f,\d\f)=\frac{1}{8\p}\int_{S_c}\bm{\epsilon}_{abcd}\d F^{cd}\d A_e\x^e\,.
\end{aligned}\ea
Combining Eq. (\ref{ftintegratedgr}) with Eq. (\ref{ftintegratedem}), the integral result of the first term in Eq. (\ref{varB}) can be expressed as
\begin{equation}
	\int_{S_c} \delta \left[\delta \boldsymbol{Q}_\xi^{\text{EM}} - \xi \cdot \boldsymbol{\Theta} \left(\phi, \delta \phi \right)  \right] = \delta^2 M - V_c \delta^2 P - \mathcal{Y} \left(\phi, \delta \phi \right).
\end{equation}
For third term in Eq. (\ref{varB}), according to the result in Eq. (\ref{lastterm2}), we can directly get
\ba\begin{aligned}
\int_\mathcal{H} \delta^2 \boldsymbol{C}_\xi = - \delta^2 \left(\int_\mathcal{H} \tilde{\boldsymbol{\epsilon}} \mu (v, r_h) T_{ae} \xi^a (dr)^e \right) - \Phi_\mathcal{H} \delta^2 Q \,.
\end{aligned}\nn\\\ea
For the fourth term of Eq. \eq{varB}, according to Eq. \eq{Cx}, we can further obtain
\ba\begin{aligned}
\int_{\S_1}\d^2\bm{C}_\x&=\int_{r_h}^{r_c} \frac{r^2\d^2 \L}{2}dr=(V_c-V_H)\d^2 P\,.
\end{aligned}\ea
Analogous to the calculation in Appendix A, for the integrated of the fifth term, we have
\ba\begin{aligned}
\d \left(\x\.\bm{E}_\f\d\f\right)&=\frac{d}{d\l}\left[\x\.\bm{E}_\f(\l)\frac{d\f(\l)}{d\l}\right]_{\l=0}\\
&=\frac{1}{2}\x\.\bm{\epsilon}\frac{d}{d\l}\left[T^{ab}(\l)\frac{d g_{ab}(\l)}{d\l}\right]_{\l=0}\\
&=\frac{1}{16\p}\x\.\bm{\epsilon}\frac{d}{d\l}\left[\L(\l)g^{ab}(\l)\frac{d g_{ab}(\l)}{d\l}\right]_{\l=0}\,,
\end{aligned}\ea
where we have used the conditions \eq{TJl} on $\S_1$. Using Eq. (\ref{gdeltag}), we can easily see that the integrated is equal to zero. Therefore, the fifth term vanishes in Eq. \eq{varB} . The sixth term in Eq. (\ref{varB}) can be neglected directly because the Killing vector contracting with the volume element is equal to zero on the event horizon. For the seventh term in Eq. \eq{varB}, since the symplectic current can be decomposed as the part of gravity and the part of electromagnetic, the term can be written as
\begin{equation}\label{integraleh}
	\mathcal{E}_\mathcal{H} = \int_\mathcal{H} \boldsymbol{\omega}^{\text{GR}} + \int_\mathcal{H}  \boldsymbol{\omega}^{\text{EM}}.
\end{equation}
According to the expression of the metric in Eq. (\ref{merticdurnper}), the integral of the gravitational part is given as
\begin{equation}
	\begin{split}
		\int_\mathcal{H} \boldsymbol{\omega}^{\text{GR}} & = - \frac{r_h}{2} \int_{v_0}^{v_1} dv \left[\delta \mu (v, r_h) \partial_v \delta f(v, r_h)\right. \\
&\left.- \partial_v \delta \mu (v, r_h) \delta f (v, r_h) \right] \\
		& = \frac{r_h}{2} \delta \mu (v_1, r_h) \delta f (v_1, r_h) = 0.
	\end{split}
\end{equation}
In the calculating process, we have used the optimal option of the first-order approximation.

On the other hand, to calculate the integral of the electromagnetic part, we should consider the specific expression of the symplectic current of the electromagnetic firstly. Utilizing the definition, the symplectic current of the electromagnetic field can be given as
\begin{equation}\label{specexpresymcurr}
	\begin{split}
		&\omega_{abc}^{\text{EM}} =  \frac{1}{4 \pi} \boldsymbol{\epsilon}_{dabc} \left[\delta A_e \mathcal{L}_\xi \delta F^{de} -\delta F^{de} \mathcal{L}_\xi \delta A_e \right] \\
		& + \frac{1}{4 \pi} \left[\left(\mathcal{L}_\xi \delta \boldsymbol{\epsilon}_{dabc} \right) F^{de} \delta A_e - \delta \boldsymbol{\epsilon}_{dabc} F^{de} \mathcal{L}_\xi \delta A_e \right].
	\end{split}
\end{equation}
According to the expressions of the electromagnetic strength and the vector potential in Eq. (\ref{S1dsa}), the index $d$ in the volume element should contribute the component of $r$, i.e., $(dr)_d$. After contracting $(dr)_d$ with $F^{de}$, we can further obtain a relation $(dr)_d F^{de} \propto \xi^e$. Using this relation and based on the gauge condition $\xi^a \delta A_a|_\mathcal{H} = 0$, we find that the last two terms in Eq. (\ref{specexpresymcurr}) are both equal to zero. Therefore, Eq. (\ref{specexpresymcurr}) can be simplified as
\begin{equation}\label{reduspecexpresymcurr}
	\omega_{abc}^{\text{EM}} = \frac{1}{4 \pi} \mathcal{L}_\xi \left(\boldsymbol{\epsilon}_{dabc} \delta A_e \delta F^{de} \right) - \frac{1}{2 \pi} \boldsymbol{\epsilon}_{dabc} \delta F^{de} \mathcal{L}_{\xi} \delta A_e.
\end{equation}
The first term on the right-hand side of Eq. (\ref{reduspecexpresymcurr}) does not appear in the integral because it only involves a boundary term which will not contribute to $\mathcal{E}_\mathcal{H}$. Therefore, the expression of Eq. (\ref{integraleh}) can be finally written as
\begin{equation}\label{intsymcurele}
	\begin{split}
		\mathcal{E}_\mathcal{H} & = \frac{1}{2 \pi} \int_\mathcal{H} \tilde{\boldsymbol{\epsilon}} (dr)_d \delta F^{de} \mathcal{L}_\xi \delta A_e \\
		& = \frac{1}{2 \pi} \int_\mathcal{H} \left[\tilde{\boldsymbol{\epsilon}} (dr)_a \xi^b \delta F^{ac} \delta F_{bc} + (dr)_a \delta F^{ac} \nabla_c \left(\xi^b \delta A_b \right) \right] \\
		& = \frac{1}{2 \pi} \int_\mathcal{H} \tilde{\boldsymbol{\epsilon}} (dr)_a \xi^b \delta F^{ac} \delta F_{bc}.
	\end{split}
\end{equation}
In the last step, we have used the gauge condition $\xi^a \delta A_a |_\mathcal{H} = 0$. According to the stress-energy tensor of the electromagnetic, the result of Eq. (\ref{intsymcurele}) can be rewritten as
\begin{equation}
	\mathcal{E}_\mathcal{H} = \delta^2 \left( \int_\mathcal{H} \tilde{\boldsymbol{\epsilon}} \mu (v, r_h) T_{ab}^{\text{EM}} (dr)^a \xi^b \right).
\end{equation}

Finally, combining above results, the second-order variational identity can be expressed as
\ba\begin{aligned}\label{Var2B}
\d^2M&-\F_H\d^2Q-V_H\d^2P=\math{E}_{\S_1}(\f,\d\f)+\math{Y}(\f,\d\f)\\
&+ \delta^2 \left[ \int_\mathcal{H} \tilde{\boldsymbol{\epsilon}} \mu (v, r_h) \left(T_{ab}^{\text{EM}} + T_{ab}\right) (dr)^a \xi^b \right].
\end{aligned}\ea

\section*{Appendix C: Derivation the expressions of $\math{E}_{\S}(\f,\d\f)$ and $\math{Y}(\f,\d\f)$}

In this appendix, we would like to derive the last remanent terms in the second-order variation identity \eq{Var2}, i.e., $\math{E}_{\S}(\f,\d\f)$ and $\math{Y}(\f,\d\f)$. Since the two terms only dependent on the first-order variation of the dynamical fields and the integral just completes on the hypersurface $\Sigma_1$, we can try to construct an auxiliary spacetime to calculate them following the trick introduced in \cite{Sorce:2017dst}. According to the stability condition, the spacetime on $\Sigma_1$ still belongs to the class of RN-AdS solutions, and the configuration of dynamical fields varying with the matter fields perturbation can also be described by the one parameter family $\lambda$. Therefore, the metric and electromagnetic strength for the auxiliary spacetime can be written as
\ba\begin{aligned}\label{dsaRA}
ds^{2}_\text{RA}(\lambda) &=-f^\text{RA}(r, \lambda) d v^{2} + 2 dv dr +r^{2}\left(d \theta^{2}+\sin ^{2} \theta d \phi^{2}\right), \\
	\boldsymbol{F}(\lambda) &=\frac{Q^\text{RA}(\lambda)}{r^2}dr\wedge d t,
\end{aligned}\ea
with the blackening factor
\begin{equation}\label{blackening3}
	f^\text{RA}(r, \lambda)=1-\frac{2 M^\text{RA}(\lambda)}{r}+\frac{[Q^\text{RA}(\lambda)]^{2}}{r^{2}}-\frac{\L^\text{RA}(\l) r^{2}}{3}\,.
\end{equation}
However, we assume that the changing behavior of the dynamical fields are only described using the first-order variation in the auxiliary spacetime, and the higher-order variation does not exist. Following this perspective, the parameters, $M^{\text{RA}}(\lambda)$, $Q^{\text{RA}}(\lambda)$, and $\Lambda^{\text{RA}}(\lambda)$ can be expanded as
\begin{equation}\label{variation2}
\begin{split}
&M^\text{RA}(\lambda)=M+\lambda \delta M\,, \\
&Q^\text{RA}(\lambda)=Q+\lambda \delta Q\,, \\
&\Lambda^\text{RA}(\lambda)=\Lambda+\lambda \delta \Lambda\,,
\end{split}
\end{equation}
where $\delta M$, $\delta Q$ and $\delta \Lambda$ are chosen to agree with the value of the first-order approximation of the perturbation. Since the auxiliary spacetime just contains the first-order variation of the dynamical fields, the relation $\d^2 M^\text{RA}=\d^2 Q^\text{RA}=\d^2\L^\text{RA} = \mathcal{E}_\mathcal{H} \left(\phi, \delta \phi \right) = 0$ can be given naturally. It is not difficult for us to check $\d \f^\text{RA}=\d \f$ on the hypersurface $\S_1$ which implies that $\math{E}_{\S_1}(\f,\d\f)=\math{E}_{\S_1}(\f,\d\f^\text{RA})$ and $\math{Y}(\f,\d\f)=\math{Y}(\f,\d\f^\text{RA})$. Therefore, $\math{E}_{\S_1}(\f,\d\f)$  and $\math{Y}(\f,\d\f)$ can be straightly derived in this auxiliary spacetime.

Integrating Eq. (\ref{var2}) on $\S_1$ and using Stoke's theorem, we can obtain
\ba\begin{aligned}\label{varC}
&\int_{S_c}\d \left[\d \bm{Q}_\x^\text{RA}-\x\.\bm{\Q}(\f^\text{RA},\d\f^\text{RA})\right]\\
&-\int_{B_1}\d \left[\d \bm{Q}_\x^\text{RA}-\x\.\bm{\Q}(\f^\text{RA},\d\f^\text{RA})\right]\\
&-\math{E}_{\S_1}(\f,\d\f^\text{RA})+\int_{\S_1}\d^2\bm{C}_\x^\text{RA}+\int_{\S_1}\d\left(\x\.\bm{E}_\f^\text{RA}\d\f^\text{RA}\right)=0\,.
\end{aligned}\ea
Analogous to the calculating process in Appendix B, and using the above relation, the last two terms vanish, and the first term becomes
\ba\begin{aligned}
\int_{S_c}\d \left[\d \bm{Q}_\x^\text{RA}-\x\.\bm{\Q}(\f^\text{RA},\d\f^\text{RA})\right]=-\math{Y}(\f,\d \f^\text{RA})\,.
\end{aligned}\ea
So Eq. \eq{varC} reduces to
\ba\begin{aligned}\label{EY}
&\math{E}_{\S_1}(\f,\d \f^\text{RA})+\math{Y}(\f,\d \f^\text{RA})\\
&=-\int_{B_1}\d \left[\d \bm{Q}_\x^\text{RA}-\x\.\bm{\Q}(\f^\text{RA},\d\f^\text{RA})\right]\,.
\end{aligned}\ea
According to Eqs. \eq{EMvar2} and \eq{GRvar2} and considering the gauge condition $\x^a \d A_a=0$ on $\math{H}$, the integral in Eq. \eq{EY} can be given as
\ba\begin{aligned}
\math{E}_{\S_1}(\f,\d \f^\text{RA})+\math{Y}(\f,\d \f^\text{RA})&=\frac{1}{4\p}\int_{B_1}\frac{\d Q^2}{r} \sin\q d\q d\f\\
&=\frac{\d Q^2}{r_h}.
\end{aligned}\ea

\end{document}